# Spatio-Temporal Mobility Patterns of On-demand Ride-hailing Service Users

Jiechao Zhang[1], Samiul Hasan[1*], Xuedong Yan[2], and Xiaobing Liu[2]

[1]Department of Civil, Environmental, and Construction Engineering, University of Central Florida, Orlando, Florida, USA
[2]School of Traffic and Transportation, Beijing Jiaotong University, Beijing, China
[*]Corresponding author email: samiul.hasan@ucf.edu

**Abstract**

Understanding individual mobility behavior is critical for modeling urban transportation. It provides deeper insights on the generative mechanisms of human movements. Different types of emerging data sources such as mobile phone call detail records, social media posts, GPS observations, and smart card transactions have been used before to reveal individual mobility behavior. In this paper, we report the spatio-temporal mobility behaviors using large-scale data collected from a ride-hailing service platform. Based on passenger-level travel information, we develop an algorithm to identify users' visited places and the category of those places. To characterize temporal movement patterns, we reveal the differences in trip generation characteristics between commuting and non-commuting trips and the distribution of gap time between consecutive trips. To understand spatial mobility patterns, we observe the distribution of the number of visited places and their rank, the spatial distribution of residences and workplaces, and the distributions of travel distance and travel time. Our analysis highlights the differences in mobility patterns of the users of ride-hailing services, compared to the findings of existing mobility studies based on other data sources. It shows the potential of developing high-resolution individual-level mobility models that can predict the demand of emerging mobility services with high fidelity and accuracy.

## INTRODUCTION

Spatio-temporal patterns of human mobility gives information on how a city functions. Understanding individual mobility behavior, from different perspectives, is important to solve many city problems such as urban planning (Sun et al., 2016, Tian et al., 2010), traffic management (Chen et al., 2016), public safety (Horanont et al., 2013, Lu et al., 2012), intelligent transportation system (Zhang et al., 2011), smart cities (Pan et al., 2013), public transportation (Zhao et al., 2018), disease spread and control (Bajardi et al., 2011, Wesolowski et al., 2012) and emerging issues such as autonomous vehicle operations (Bansal and Kockelman, 2017) and mobility as a service design (Jittrapirom et al., 2017). In recent years, a wide range of emerging human movement data sources—such as bank notes (Brockmann et al., 2006), social media data (Hasan et al., 2013b, Jurdak et al., 2015), mobile phone call detail records (Huang et al., 2018), smart card transactions (Zhao et al., 2018), and floating car observations (Chen et al., 2019, Peng et al., 2012, Veloso et al., 2011, Zheng et al., 2018)—have been used to uncover individual mobility behavior (Gonzalez et al., 2008) and commuting patterns (Ma et al., 2017). In this paper, we report the findings on the mobility behavior of a new population group—users of emerging on-demand ride-hailing services—after analyzing large-scale trip data from a ride-hailing platform.

With the emergence of ride-hailing services such as Uber, Lyft, and Didi, massive passenger movement data from these platforms have a tremendous potential to reveal individual travel behavior patterns. In this study, we analyze the spatio-temporal patterns of individual mobility using the movement data extracted from Didi, a Chinese ride-hailing service. First, we present a distance-based algorithm to identify the visited places of different passengers. Second, given the visited places of passengers, we investigate the spatio-temporal patterns of individual movements. Then, for every individual user, we detect their home and work place based on the probability of visiting different places at different time periods (morning and evening peak hours). Finally, we reveal individual mobility patterns when using ride-hailing services from different perspectives such as trip generation, gap time, number of visited places and their rank, spatial distribution of home and work place, travel distance, and travel time for both commuting and non-commuting trips. The resulting distributions show the potential of modeling the generative mechanism of ride-hailing service demand. Such models will enable high-fidelity (e.g., individual level) simulation of demand prediction, dispatching, ride sharing, and pricing applications of ride-hailing services.

To the best of our knowledge, this is the first study that reveals individual-level mobility patterns of ride-hailing service users based on large-scale data available from a ride-hailing platform. The main contributions of this paper are as follows:

- We reveal the spatio-temporal mobility patterns of ride-hailing service users. Although ride-hailing platforms have been serving demand for several years, previous studies did not investigate the mobility patterns of the users of these services.

- We investigate critical aspects of on-demand ride-hailing services such as the gap time between two consecutive rides and the rank of visited places of the users.

- We fit the distributions of travel distance and travel time of on-demand ride-hailing service trips. The parameters of these distributions can be used to establish trip generation mechanisms for agent-based simulations which will significantly benefit the operations and management of on-demand ride-hailing services as well as urban planning and traffic management.

This paper is organized as follows. Section 2 describes the study area and the dataset extracted from Didi's platform. Then, a heuristic algorithm based on distance is presented to detect the most visited places of ride-hailing service users. Section 3 reports the empirical results observed from individual mobility patterns including the distribution of generated trips in different time periods both for commuting and non-commuting purpose, rank of the visited places, gap time between two consecutive trips, the spatial distribution of home and work place and the general distribution of travel distance and travel time. Section 4 discusses the results of individual mobility patterns, compares the results with previous studies, and provides the insights and implications for transportation planning and operations. Section 5 concludes with the limitations of this research that can be improved in future work.

**Literature Review**

Individual mobility behavior reflects the spatio-temporal dynamics of urban mobility at a high resolution (Brockmann et al., 2006). Traditionally, survey based travel data have been used to analyze and model individual mobility behavior. Although travel survey methods have evolved from traditional pen-and-paper based data collection to nowadays web and smartphone-based data collection (Wolf et al., 2001) approaches. High cost and low sample size are major challenges towards implementing these tools at scale (Wu et al. 2011). To overcome the limitations of travel surveys, researchers have analyzed emerging data sources such as bank notes (Brockmann et al., 2006), mobile phones (Gonzalez et al., 2008), and social media (Rashidi et al., 2017) data for understanding and modeling individual mobility behavior.

With widespread adoption of mobile phones and location-based services, various large-scale high-resolution datasets with varying capabilities have been used to understand individual mobility behavior (Alessandretti et al., 2017, Zhang et al., 2018). For instance, call detail records (CDR) from mobile phones can provide useful insights on individual mobility at a scale that was unimaginable before (Gonzalez et al., 2008, Chen et al., 2016, Huang et al., 2018). However, CDR data are generated when a person makes a phone call or sends a message. It is a challenging task to predict when and where an individual will use his/her phone, which may result into incomplete travel information. Thus, inferring the origin and destination of individual activity is difficult based on such data (Huang et al., 2018). Social media posts can also provide rich information on individual travel and activity behavior (Rashidi et al., 2017). Through mining geo-location data recorded when user's check-in or post in social media platforms, individual activities can be identified over a long period, offering useful insights on individual travel patterns. However, these data do not include the precise start and end time of a trip, limiting applications in transportation (Hasan and Ukkusuri, 2018). Data from social media and mobile phones are defined as extrinsic mobility data that do not directly observe individual travel behavior (Zhao et al., 2018).

Different from extrinsic mobility data (Zhao et al., 2018), smart card data (Hasan et al., 2013b) and floating car data (FCD) (Ehmke et al., 2012) can be defined as intrinsic mobility data that are directly collected from transportation system operations. For instance, smart card data are extracted from public transit operations, while FCD are collected from taxicab. Both types of datasets record when and where a user takes public transit (e.g., subway or bus) or taxi for a trip—giving precise information on the origin, destination, distance, price, and time of a trip. Unlike extrinsic data, intrinsic mobility data can offer mode-specific complete trajectory information, giving a different perspective to understand individual travel behavior.

Compared to smart card data, taxicab data have limitations to uncover individual mobility patterns, because passengers always pay in cash or credit cards when they take a taxi, without requiring the system to record individual details for historical tracking. Thus, previous studies on human mobility using taxicab data focused on urban resident's aggregate travel patterns (Zheng et al., 2018) or taxi drivers' travel behavior (Leng et al., 2016), instead of analyzing individual passenger's mobility patterns. Due to the lack of available data, studies on individual mobility patterns using taxicab services hardly exist. However, a deeper understanding of individual mobility patterns under taxicab services from a passenger's perspective is significantly beneficial to many problems involving emerging ride-hailing services such as real-time demand prediction (Ke et al., 2017), designing ride-sharing operations (Alonso-Mora et al., 2017a), and designing mobility services for autonomous vehicles (Alonso-Mora et al., 2017b).

The emergence of on-demand ride-hailing platforms provides an innovative transportation service that can be easily requested via a smartphone app—providing longitudinal mobility data at an individual level (Contreras and Paz, 2018). These ride-hailing service platforms have a great potential in revealing individual mobility behaviors since the locations and timings of individual trips can be recorded through the GPS devices in the smartphone and stored in the platform. However, previous studies mainly used ride-hailing data for analyzing aggregate mobility behavior (Dong et al., 2018) and solving traffic modeling and prediction problems (Ke et al., 2017). As such, human mobility literature lacks an understanding of individual-level mobility patterns based on ride-hailing service data. To fill this research gap, we use large-scale data extracted from a major ride-hailing platform to analyze the mobility patterns of its users. To the best of our knowledge, this is the first research that presents individual mobility behavior from on-demand ride-hailing service data. The results of this research will provide valuable insights for many future studies such as demand prediction, policy making, and design of ride-hailing services.

## DATA AND METHODS

### Study Area and Data Description

In this study, we have analyzed a large-scale dataset from Didi (the largest ride-hailing service company operating in Beijing, China). The study region covers the area inside Beijing's 6$^{th}$ ring road, seen as **Figure 1**. The dataset used in this paper was extracted from Didi from March 1, 2017 to June 31, 2017. The dataset records more than 3 million Didi users with around 200 million trips. **Table 1** presents the fields available in the data with their description. To find the trip records of a specific passenger, we have utilized the ID (P_id) of a passenger used in Didi. In this study, we have randomly selected 50,000 users who have made more than 10 trips per month in the data collection period.

**TABLE 1 Detailed Data Attributes**

| Fields | Field Name | Field Type | Field Description |
|---|---|---|---|
| R_id | Record ID | String | The record id of one trip |
| P_id | Passenger ID | String | The passenger encrypted id of one trip |
| D_id | Driver ID | String | The driver encrypted id of one trip |
| O_LNG | Longitude of Origin | Floating | The longitude of the origin |
| O_LAT | Latitude of Origin | Floating | The latitude of the origin |
| D_LNG | Longitude of Destination | Floating | The longitude of the destination |
| D_LAT | Latitude of Destination | Floating | The latitude of the destination |
| O_Time | Start Time | Timestamp | The timestamp of the origin |
| D_Time | Arrive Time | Timestamp | The timestamp of the destination |
| L | Travel Distance | Floating | The travel distance of the trip |

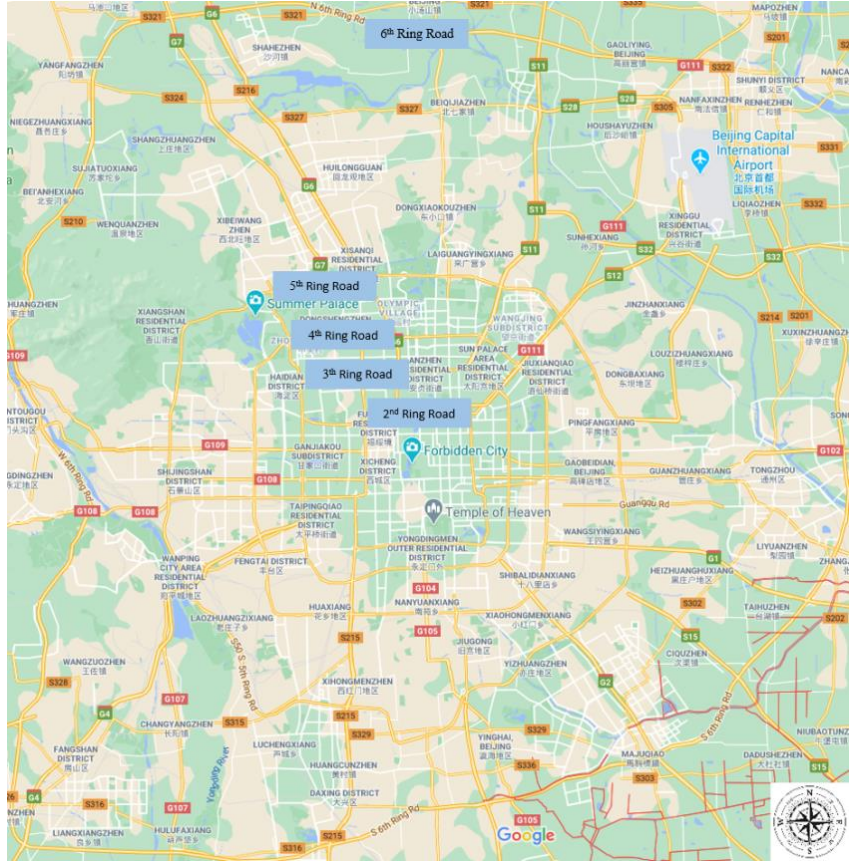

**Figure 1 The study region: area inside Beijing 6th ring road**

Raw movement data from a ride-hailing platform have several issues. For example, GPS errors may be caused by either blockage of the GPS signal or hardware/software bugs during the data collection process. To clean the raw data, we followed the following steps:

*Step 1:* Convert the current coordinate system of Didi's data to the Worldwide Geodetic System 1984 (WGS84) coordinate system;
*Step 2:* Remove the data which have the coordinates (origin or destination) outside Beijing's 6th ring road;
*Step 3:* Based on the speed limitation, remove the trips with average travel speed ($Distance/(D\_Time - O\_Time)$) above 120 km/h.

**Distance-Based Visited Place Generation Algorithm**

Passenger movement data provide the GPS coordinates of origins and destinations. However, two origins or destinations can belong to the same place (e.g., home or work place) with different coordinates possibly due to different boarding points from the same location. Based on an individual's origins and destinations, her home and workplace can be inferred. Passengers typically leave home and arrive at workplace in the morning peak hour and have a reverse travel direction in the evening peak hours. However, for some commuters using ride-hailing service, they only use ride-hailing service to travel to either residence or workplace and take another traffic mode for the other. There are also some users do not utilize the ride-hailing service for commuting purpose. Thus, to identify the function (home or workplace) of visited places, we should consider the ratio of the number of trips in peak hours to the total trip number for each individual.

In this study, we have applied several heuristic rules to identify the visited places and their functions for each individual as follows:

***Rule1:*** For each individual user, if the distance between two locations (origins or destinations) is less than 500m, then these two locations are defined as the same place.

***Rule2:*** For each individual user, for each location, we count the number of trips originated from the same location in the extended morning peak hour (6 am – 11 am) or the number of trips ended at the same location in the extended evening peak hours (3 pm – 8 pm), among all the trips made by the user in the analysis period. If the ratio of the largest count to the total trip number is more than 40%, then, the location with the largest count is defined as the home place. Otherwise, the individual's home place cannot be identified.

***Rule3:*** For each individual user, we count the sum of the number of destinations in the extended morning peak hours (6 am – 11 am) and the number of origins in the extended evening peak hours (3 pm – 8 pm) for each location. If the ratio of the largest count to the total trip number is more than 40%, then, the location with the largest count is defined as the workplace. Otherwise, the individual's workplace cannot be identified.

According to ***Rule1***, we developed a distance-based visited place generation algorithm (*DBVPGA*) to identify the visited places of each individual user. According to the distance between different coordinates, the *DBVPGA* will detect whether the origin or destination is a new place and then assign an ID, as a visited place (start from 0), to the origin and/or destination. The key definitions of *DBVPGA* are shown as follows:

- $d_{th}$: the threshold distance used to identify places ($d_{th}$ = 500 meters).
- $d_{i,j}$: the distance between point $i$ and point $j$.
- $O_n$: the $n^{th}$ origin point of a user.
- $D_n$: the $n^{th}$ destination point of a user.
- *MTN*: the number of trips per month made by a user.
- *PID*: the ID of visited places of a user (0, 1, …, n).
- *maxPID*: the maximum PID of a user.
- *VPF*: the types of visited places of a user (0-home, 1-work, 2-other).

We develop an algorithm to convert the coordinates of the origins and destinations into relative IDs of visited places (*PIDs*) for an individual. Briefly speaking, for each individual user, we have a list of coordinates of origins and destinations of all the trips made by the user. Every origin and destination are defined as different points and we start from the origin of the first trip by setting it as the first visited place (i.e., *PID* = 0). Then, choose the destination of that trip as the second point and compare the distance between this point and the previous point ($d_{i,j}$) with the threshold distance ($d_{th}$). If $d_{i,j}$ is more than $d_{th}$, then set the second point as a new visited place, the *PID* of the second place is 1 and the *maxPID* is added by 1. In this way, two different visited places are generated. If $d_{i,j}$ is less than $d_{th}$, then the second point is the same visited place as the first point, and the *PID* of the second place is also 0. Likewise, for the other points, we compare the distance between them and existing visited places with $d_{th}$, if all the distances $d_{i,j}$ are more than $d_{th}$, then generate a new visited place and the *maxPID* is added by 1. Otherwise, if the distance between the point and any existing place is less than $d_{th}$, then the *PID* of this point will be the same as the specific existing place. We iterate over all the points until every point has its own *PID*. The algorithm is described as follows:

---

**Algorithm 1** Distance-Based Visited Place Generation Algorithm

**Input:** Passenger ID, the list of coordinates of origins and destinations, number of trips per month (MTN)

**Output:** PID

---

1. For each individual user:
2.     $PID_{o1} = 0$, maxPID = 0.
3.     For i from 2 to MTN:
4.         ON = i – 1
5.         For j from 1 to ON:
6.             If $l_{oi,oj} < l_{th}$:
7.                 $PID_{oi} = PID_{oj}$
8.             End if
9.         End for
10.        If $PID_{oi}$ = N/A:
11.            maxPID = maxPID +1
12.            $PID_{oi}$ = maxPID
13.        End if
14.     End for
15.     For i from 1 to MTN:
16.         If $l_{di,oi} < l_{th}$:
17.         $PID_{di} = PID_{oi}$
18.         End if
19.         DN = i – 1
20.         For j from 1 to DN:
21.             If $l_{di,dj} < l_{th}$:
22.                 $PID_{di} = PID_{dj}$
23.         End for
24.         If $PID_{di}$ = N/A:
25.            maxPID = maxPID +1
26.            $PID_{di}$ = maxPID
27.         End if
28.     End for
29. End for

---

After running the algorithm, the visited places of each user are generated, the function (home place or work place) of the visited places can be identified according to ***Rule 2*** and ***Rule 3***.

**Fitness metrics**

In this paper, to reveal the statistical distributions of individual travel distance and travel time, we apply several commonly used distributions to the corresponding trip attributes for the selected users, including *log-normal*, *Weibull*, *gamma*, *student's t*, *exponentiated Weibull* and *power log-normal* (see **Appendix** for more information on these distributions).

The parameters of the distributions are estimated by maximum likelihood methods and the detailed information can be found in literature (Myung, 2003). Besides, we use Kolmogorov–Smirnov test (K-S test) (Massey Jr, 1951) to evaluate the performance of the fitness. The *null hypothesis* of the K-S test is that the two distributions are identical.

The empirical distribution function $F_n$ for *n* independent and identically distributed (iid) ordered observation $X_i$ is defined as:

$$F_n(x) = \frac{1}{n}\sum_{i=1}^{n} I_{[-\infty,x]}(X_i) \qquad (1)$$

Where $I_{[-\infty,x]}(X_i)$ is the indicator function, equal to 1 if $X_i \leq x$ and equal to 0 otherwise. The Kolmogorov–Smirnov statistic for a given cumulative distribution function *F(x)* is:

$$D_n = sup_x |F_n(x) - F(x)| \qquad (2)$$

where $sup_x$ is the supremum of the set of distances.

**EMPIRICAL RESULTS**

**Temporal Pattern - Trip Generation**

Based on the origin and destination information, a trip can be characterized by its travel purpose. In this study, we analyze the individual trip generation patterns by decomposing the on-demand ride-hailing service trips into two groups – commuting and non-commuting trips according to their travel locations. We define a trip which is made between the residence and the work place of a user as a commuting trip. When a trip contains at least one location which is neither the residence nor the work place, we define it as a non-commuting trip.

**Figure 2a** shows the daily trip distribution of commuting and non-commuting trips of all selected users from the on-demand ride-hailing service. It reveals weekly periodicity of passengers' travel behavior. On-demand service users tend to travel more frequently on weekdays compared to weekends and festivals (May 1st, the Labor Day, is a holiday in China). However, the periodicity characteristics for the commuting and non-commuting trips show significant differences2. In terms of commuting trips, since people always work on weekdays, the number of commuting trips decreases sharply on weekends. The demand of non-commuting trips is more than that of commuting trips for ride-hailing service. The weekly periodicity indicates that the demand on weekends of non-commuting trips will increase smoothly. The total demand of commuting and non-commuting trips is higher on weekday compared to weekends.

**Figure 2b** presents the hourly distribution of trips, indicating that the on-demand ride-hailing service trips for commuting have a typically bimodal distribution while the distribution for non-commuting trips is unimodal. For commuting trips, peak demand is seen from 7 am to 9 am (morning peak hour) and from 5 pm to 9 pm (afternoon peak hour). For non-commuting trips, peak demand is seen from 5 pm to 9 pm. The highest demand of ride-hailing service is seen around 8 am for commuting purpose. From 10 am to midnight, the demand of non-commuting trips exceeds the demand of commuting trips.

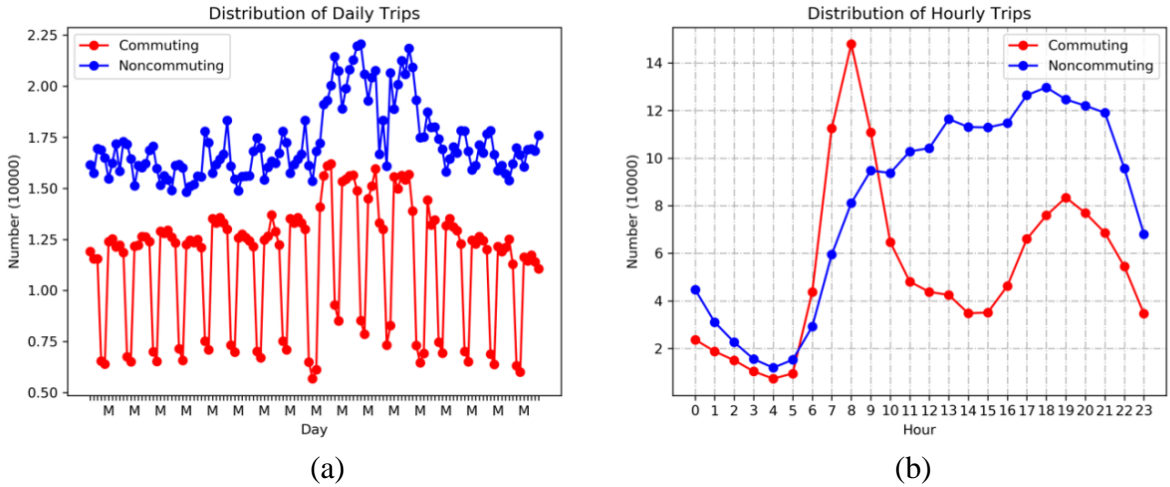

(a)                                                (b)

**Figure 2 The probability of the demand of ride-hailing service (March 1-June 30, 2017): (a) the distribution of the number of daily trips using the ride-hailing service ('M' means Monday of every week) (b) the distribution of average hourly trips number of ride-hailing service.**

**Gap Time**

One of the most important characteristics of a ride-hailing service is how long does it take for a user to make the next trip. To uncover the distribution of the time spent to make a new trip, we create a variable called the gap time. It is defined as the difference in start time between two consecutive trips for a given user. To determine this distribution, we select all the users who have made at least 2 trips in a month.

**Figure 3a** shows the distribution of average gap time of users with the number of trips made in a month. It indicates that with the increase of monthly trips, the gap time between consecutive trips decreases sharply at the beginning; the decreasing trend turns slower when the number of monthly trips is greater than 40. The maximum gap time is found as around 8,000 min (5.6 days) when users have only 2 monthly trips. When the number of monthly trips is more than 80, the gap time is close to 300 min (or 5 hours).

**Figure 3b** presents the distribution of gap time of all the users in the observation period (one month), with peaks at two-hour and one-day gaps, respectively. This finding is different from the current mobility research (Hasan et al. 2013) which shows a 9-hour peak when it comes to the public transport users. Additionally, the distribution also indicates multi-day local peaks, which might be closely related to the passengers' regularity patterns in requesting rides.

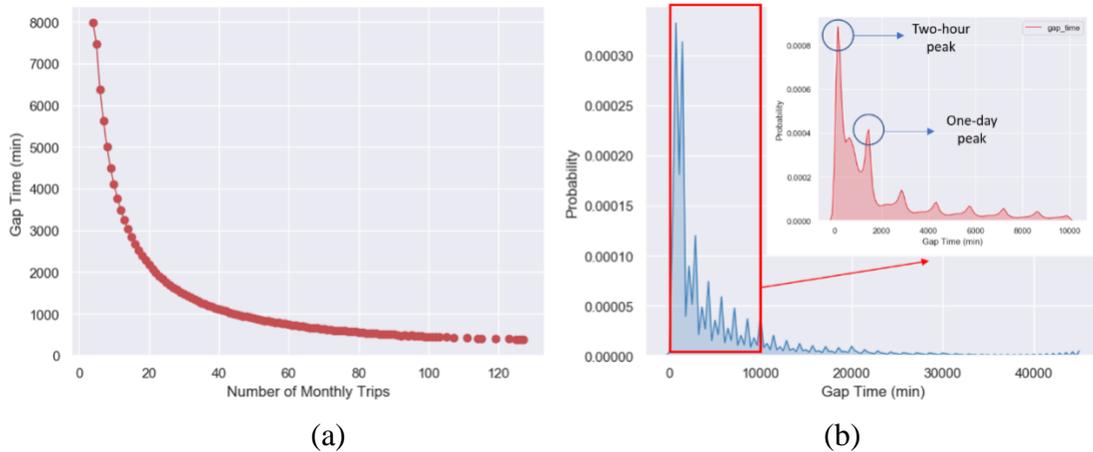

(a)                          (b)

**Figure 3 The distribution of gap time. (a) the average gap time (min) vs. their corresponding users' groups with different number of trips per month; (b) the probability of different gap time (min) in two scales: the longer range shows all the possible gap time and the inset shows the range only between 0 to 10,000 min.**

**Number of Visited Places and their Rank**

Each ride-hailing service user visits a specific number of different locations within the observation period. We rank the visited places based on the frequency of visits to those places and determine the probability of visiting each place. For instance, for a user, a place with rank 1 means the most visited location, a place with rank 2 represents the second-most visited location, and so on.

The number of visited places and the rank of those locations play an essential role for mobility pattern analysis. To uncover the distribution of the number of visited places and the rank of visited places, we select the passengers who have at least 10 monthly trips so that enough trips are generated to reveal the ranking patterns.

**Figure 4a** presents the distribution of the number of visited places—indicating that majority of the frequent users of the ride-haling service visit on average 8 to 12 different places in a month. Given that a frequent user makes at least 10 monthly trips and every trip contains two places (origin and destination), he/she has a high probability to visit the same places when using the service. To uncover the regularity patterns, the probability distribution of visiting a place over the rank of the visited place is presented in log-log scale in **Figure 4b**. From the distribution, we can find that most of the users' trips are concentrated in a few locations, especially the first rank visited place. For instance, users who visited 5 different places, the most visited place accounts for nearly 50% of the total trips. When a user visits more places, the probability of the most visited place slightly declines. Additionally, the probability of the second most visited place (i.e., rank =2) is close to the most visited place when the number of visited places is low. Users who have visited more places, the difference between the probability of the first two rank visited places becomes larger. The distribution of the visited place rank follows a Zipf's law when the number of visited places is high.

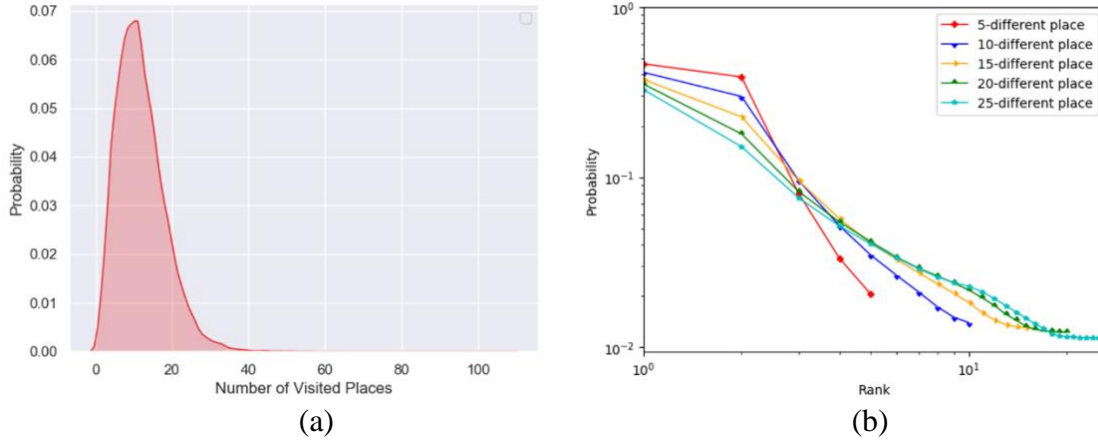

(a)                      (b)

**Figure 4 The distribution of passengers' number of visited places and the probability of visiting a place against its rank: (a) the probability distribution of the number of visited places (the number of different locations visited by a user in the observation period); (b) the probability of visiting different places against its rank in log-log scale (the legend refers to the groups visiting a given number of different places).**

**Spatial Distribution of Home and Work Place**

According to the distance-based visited place generation algorithm, we detect the home and work place of on-demand ride-hailing service users. We then visualize the heatmap of the work and home place in a map as shown in **Figure 5,** where red color means high density regions and blue color means low density regions. From the heatmap of both home and workplace, we can find that the spatial distribution of home place is more dispersed while the workplace distribution is more centralized. Majority of the dense work zones are located inside fifth ring road while many dense home zones are located outside the fifth ring road, which may lead to the imbalance in job-housing distribution. The workplaces are more concentrated on the eastern of Beijing, such as the districts of Guomao, Wangjing, Sanlitun, which are the most famous business districts. In the west side of Beijing, Zhongguancun district is the most concentrated work place region for on-demand ride-hailing users. For the home place, it indicates that most of the on-demand ride-hailing users live in districts such as Panjiayuan, Huilongguan, Pingguoyuan, and Xihongmen, which are some of the largest residential districts in Beijing.

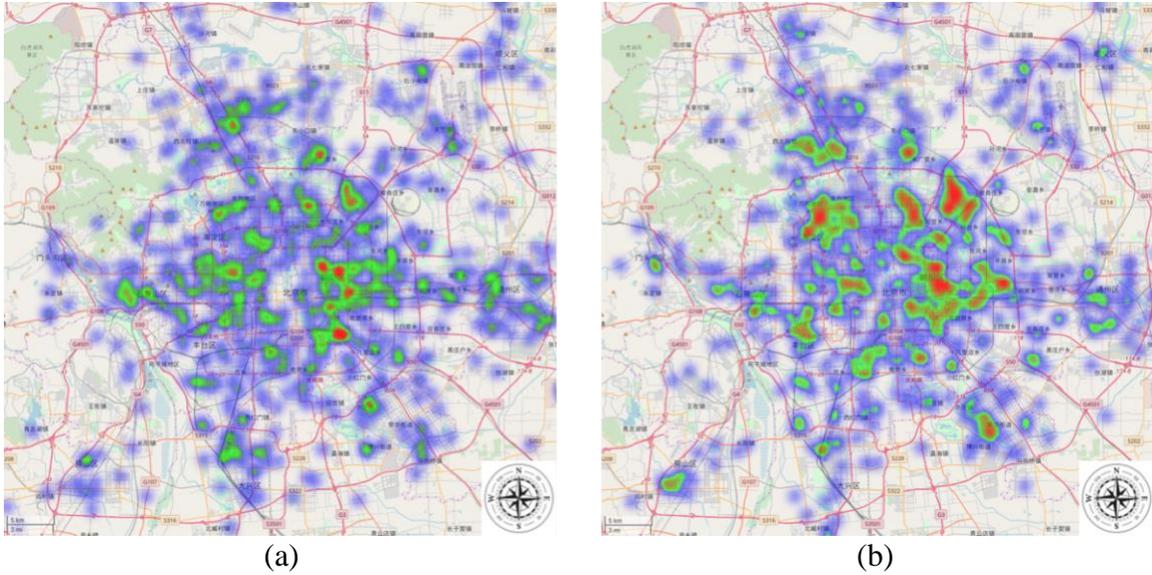

(a)                                                                 (b)

**Figure 5 The spatial distribution of passengers' home place and work place (blue color means low density and red color means high density): (a) the spatial distribution of passengers' home place; (b) the spatial distribution of passengers' work place.**

**Spatial Distribution of Travel Distance**

For on-demand ride-hailing service, the average travel distance per trip is heterogeneous from the perspective of travel purpose (commuting or non-commuting) and spatial scale. To reveal the heterogeneity in travel distance of ride-hailing service users, we divide the study area into 30X30 grids with a grid size of about 2X2km$^2$. The grids are identified by indices from left to right (horizontally from 0 to 29) and from top to bottom (vertically from 0 to 29), as shown in **Figure 6**. Then, we assign each trip to a grid based on its origin coordinate and aggregate all the trips originating from each grid. Finally, we calculate the average travel distance of the trips which are aggregated over each grid and visualize the spatial distribution of travel distance in a map.

**Figure 6** presents the results of the spatial distribution of travel distance in different regions in the study area for both commuting and non-commuting trips. In general, we observe that the travel distance of non-commuting trips is more than that of commuting trips. Users outside the 5$^{th}$ ring have longer commuting trips and users inside the 5$^{th}$ ring road make commuting trips shorter than 8 km. The spatial distribution of non-commuting trips from **Figure 6b** also shows interesting characteristics. Users from most of the regions make non-commuting trips longer than 8 km. In particular, users from regions outside the core area make trips longer than 12 km. The travel distances of the trips originated from the airport region appear different in the two distributions shown in **Figure 6**. Since a lower number of users depart from airport region for commuting purpose, the average travel distance of the commuting trips from airport region (lower than 8 km) is significantly lower than the average travel distance of non-commuting trips (greater than 20 km).

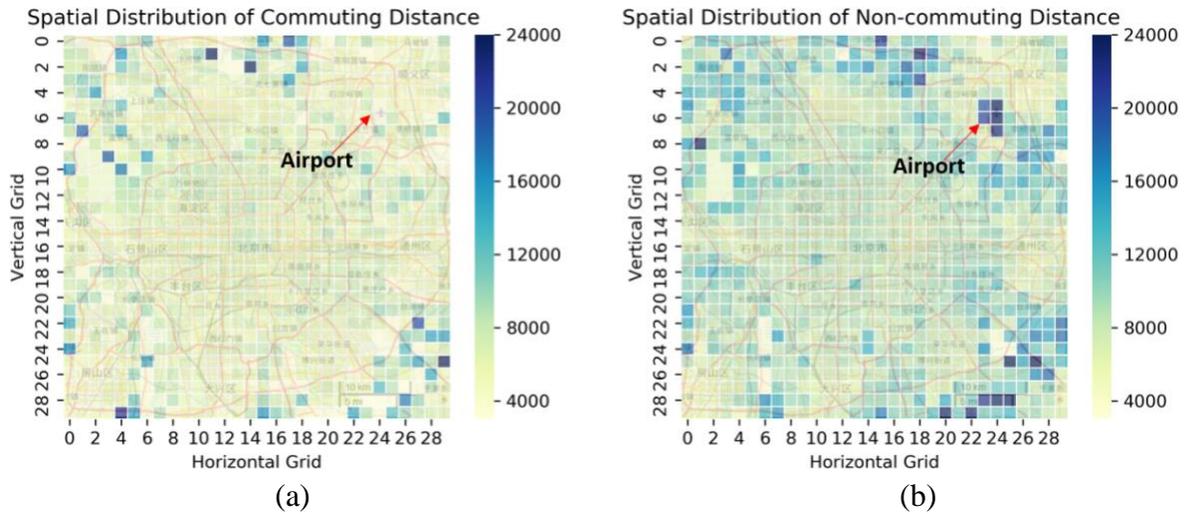

(a)                                           (b)

**Figure 6 The distribution of average travel distance in spatial scale, in which the horizontal grid means the ordered grid id in horizontal level and the vertical grid means the ordered grid id in vertical level: (a) the spatial distribution of commuting trips' average travel distance aggregated in different grids; (b) the spatial distribution of non-commuting trips' average travel distance aggregated in different grids**

**Distribution of Travel Distance**

To reveal the spatial patterns of individual mobility using ride-hailing service, the distributions of travel distance for both commuting and non-commuting trips are analyzed. In the existing studies, public transport smart card data and mobile phone data are commonly used for understanding individual mobility patterns. These data provide approximate distances of individual movements. However, movement data extracted from a ride-hailing service platform offer us a more accurate travel distance data instead of approximate displacements (Wang et al. 2015), since the locations of each trip's origin and destination can be accurately determined. Additionally, features of commuting and non-commuting trips hold significant information of urban travel behavior, which are seldom investigated due to data source limitation. To uncover individual commuting and non-commuting travel behavior, we choose the frequent users (making more than 10 trips per month). In addition, we also fit the travel distance distribution by 6 commonly used distributions mentioned before, which will be beneficial for establishing a generative mechanism to simulate the demand of ride-hailing services.

**Figure 7** shows the distributions of travel distance of commuting and non-commuting trips. Similar to **Figure 6**, we also find here that the average travel distance of non-commuting trips is more than that of commuting trips. The peaks of the two distributions occur at around 5 km. The average travel distance of commuting trips and non-commuting trips are 6.298 km and 8.467 km, respectively. Trips longer than 15 km account for 5.58% and 14.62% of the commuting and non-commuting trips, respectively. This is expected as people do not prefer to make long commuting trips by taxicab or on-demand ride-hailing service and previous studies (Wang et al. 2015) also found similar results.

To capture travel distance distribution of the on-demand ride-hailing service, we use 6 statistical distributions - *log-normal, Weibull, gamma, student's t, exponentiated Weibull and power log-normal* – to fit the travel distance distribution with a K-S test to evaluate the performance. **Table 2** presents the results of K-S test for travel distance. From the results, we can find that power log-normal distribution fits best for both the commuting travel distance and non-commuting travel distance. The power log-normal distribution has a lower D value (0.235 for commuting trips and 0.092 for non-commuting trips) and higher p-value (0.104 for commuting trips and 0.977 for non-commuting trips), which indicates a higher probability to accept the null hypothesis that the two distributions come from one identical distribution.

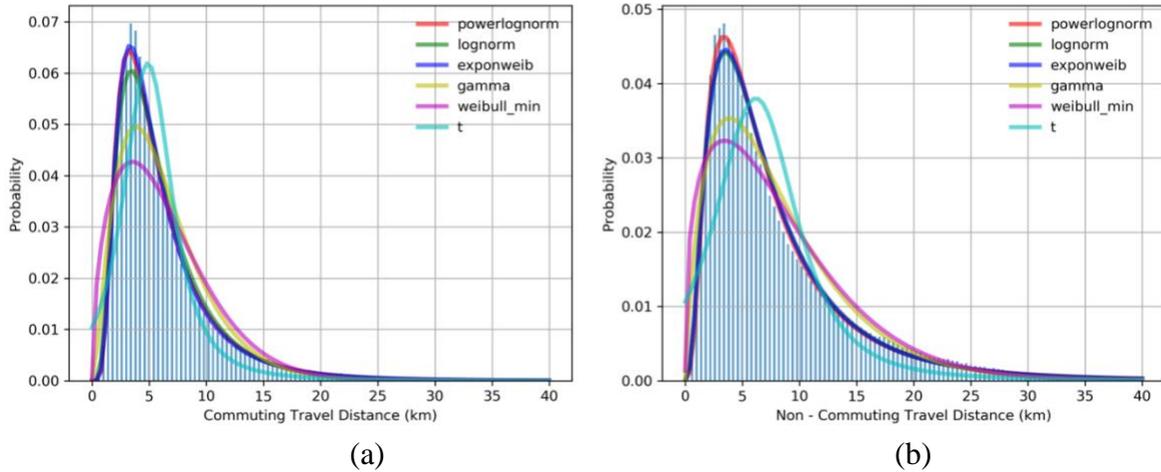

(a)          (b)

**Figure 7 The distributions of travel distance per trip with fitting curves of selected distributions: (a) distribution of travel distance (km) per trip for commuting trips; (b) distribution of travel distance (km) per trip for non-commuting trips.**

**Table 2 The results of K-S tests of selected distributions of travel distance**

| Distribution K-S Test | Commuting Distance | | | Non-commuting Distance | | |
|---|---|---|---|---|---|---|
| | D | *p*-value | Parameters | D | *p*-value | Parameters |
| Power log-normal | **0.235** | **0.104** | $p = 5.64; \sigma = 1.00$ | **0.092** | **0.977** | $p = 0.95; \sigma = 0.74$ |
| Log-normal | 0.255 | 0.062 | $\mu = 1.52; \sigma = 0.61$ | 0.112 | 0.875 | $\mu = 1.85; \sigma = 0.75$ |
| Exponential Weibull | 0.255 | 0.062 | $k = 6.01; \alpha = 0.89; \lambda = 2.54$ | 0.112 | 0.875 | $k = 13.27; \alpha = 0.50; \lambda = 0.72$ |
| Gamma | 0.608 | 0.000 | $\alpha = 1.04; \beta = 1.85$ | 0.176 | 0.377 | $\alpha = 1.88; \beta = 4.39$ |
| Weibull | 0.314 | 0.010 | $k = 1.53; \lambda = 7.02$ | 0.275 | 0.036 | $k = 0.80; \lambda = 3.53$ |
| Student's t | 0.355 | 0.002 | $v = 2.13$ | 0.235 | 0.104 | $v = 2.32$ |

**Distribution of Travel Time**

Besides the travel distance, the travel time of each on-demand ride-hailing service trip also plays an important role in revealing the spatial regularity. **Figure 8** presents the distributions of travel time for both commuting and non-commuting trips. We observe that the peaks of the commuting and non-commuting trips appear at 12 min and 14 min, respectively. The average travel time for commuting and non-commuting trips are 22 min and 26 min, respectively. In general, the travel time for non-commuting trips is more than that for commuting trips, similar to the results found with respect to the distribution of travel distance. In addition, we also use 6 commonly used distribution to fit the travel time distribution, and the results can be seen as **Table 3**. The results of K-S tests show that the commuting travel time distributions fit best to the Exponential Weibull distribution (with D = 0.157 and *p*-value = 0.527) and the non-commuting travel time distributions fit best to the log-normal distribution (with D = 0.058 and *p*-value = 1.000)**.**

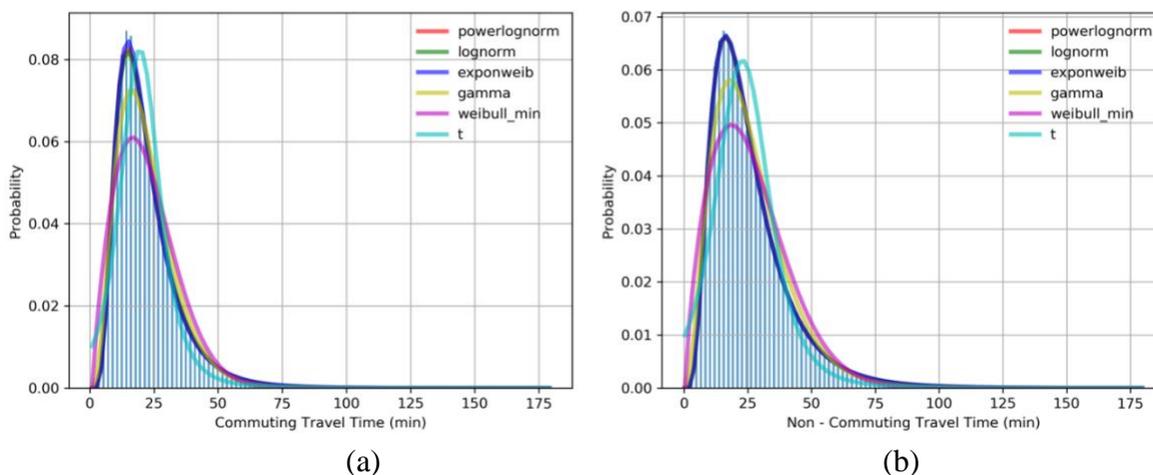

(a)                                                             (b)

**Figure 8 The distribution of travel time per trip with fitting curves of selected distributions: (a) distribution of travel time per trip for commuting trips; (b) distribution of travel time per trip for non-commuting trips.**

**Table 3 The results of K-S tests of selected distributions of travel time**

| Distribution K-S Test | Commuting Travel Time | | | Non-commuting Travel Time | | |
|---|---|---|---|---|---|---|
| | D | *p*-value | Parameters | D | *p*-value | Parameters |
| Power log-normal | 0.181 | 0.347 | $p = 0.12$; $\sigma = 0.23$ | 0.081 | 0.995 | $p = 0.15$; $\sigma = 0.22$ |
| Log-normal | 0.360 | 0.002 | $\mu = 0.62$; $\sigma = 1.52$ | **0.058** | **1.000** | $\mu = 3.16$; $\sigma = 0.52$ |
| Exponential Weibull | **0.157** | **0.527** | $k = 11.00$; $\alpha = 0.77$; $\lambda = 4.93$ | 0.083 | 0.993 | $k = 15.29$; $\alpha = 0.63$; $\lambda = 3.53$ |
| Gamma | 0.412 | 0.000 | $\alpha = 3.58$; $\beta = 5.89$ | 0.294 | 0.020 | $\alpha = 3.19$; $\beta = 8.02$ |
| Weibull | 0.471 | 0.000 | $k = 1.86$; $\lambda = 23.92$ | 0.255 | 0.062 | $k = 1.64$; $\lambda = 31.40$ |
| Student's t | 0.200 | 0.236 | $v = 3.21$ | 0.093 | 0.976 | $v = 3.69$ |

## DISCUSSION

In this study, we have analyzed large-scale trip data extracted from a ride-hailing service platform (Didi) in China to understand individual mobility patterns. To the best of our knowledge, this is the first study reporting the mobility behavior of ride-hailing service passengers. Human mobility can be characterized as the movement patterns in a spatio-temporal scale. To uncover the spatio-temporal patterns of individual movements, we have analyzed the distribution of the trip generation, gap time, number and rank of visited places, travel distance, and travel time. In addition, to capture the patterns of commuting behavior, we divide the trips into two groups: commuting and non-commuting trips according to the travel purpose of individuals.

      For temporal patterns, first, the distributions of daily and hourly trips reveal the regularities of trips made by ride-hailing services. It reveals that people tend to use on-demand service more on weekdays with 20% less trips on weekends. Additionally, the patterns of hourly trip generation distributions show differences between commuting and non-commuting trips. The distributions of trip generation for commuting trips reveal a bimodal distribution and a unimodal distribution for non-commuting trips. The morning peak hours of non-commuting trips vanish because most of

non-commuting trips are for leisure or entertainment activities. Previous study (Ma et al., 2017) analyzing public transit data, however, found that trip generation distributions of commuters and non-commuters have both morning and afternoon peak hours. The total number of commuting trips are significantly greater than that the number of non-commuting trips. This result differs with our results indicating that compared to public transit, on-demand ride-services are less preferred by commuters.

Another important aspect reported in this paper is the distribution of gap time between consecutive ride-hailing trips. In recent years, most studies have analyzed the waiting time or stay time patterns based on mobile phone data (Gonzalez et al., 2008) and interval distribution of taxi trajectory from drivers' perspective (Veloso et al., 2011, Zheng et al., 2018). No study has investigated the distribution of the time interval between two consecutive trips of ride-hailing service users. To fill this gap, we have analyzed the average gap time (time interval between two consecutive trips) distributions of on-demand service users from two aspects. First, the amount of gap time between consecutive trips is inversely proportional with the number of monthly trips. Second, the distribution of gap time follows a log-normal distribution with local spikes, similar to the patterns observed for stay time distributions from smart card data (Hasan et al., 2013a) and the return time based on mobile phone data (Gonzalez et al., 2008). In addition, previous research (Gonzalez et al., 2008) also found that the distribution of gap time is characterized by several local peaks at 24h, 48h, 72h, and so on--showing a strong temporal regularity inherent to human mobility. However, from the ride-hailing service users' perspective, the gap time probability has a two-hour peak compared to the 9-hour peak found in the previous studies (Hasan et al., 2013a). It probably indicates that people tend to use public transportation for commuting trips which have a 9-hour peak, while people prefer to use on-demand services for leisure or flexible activities which have a two-hour peak.

To find the spatial regularities of individual mobility, we identify each user's visited places and rank those places according to the number of times a user has visited a place. From the results, we find that frequent users tend to visit 8-10 different locations and visit the top two locations more. This shows a great regularity pattern of the on-demand service users when they make a trip. Previous research found similar results with the data extracted from smart card transactions (Hasan et al., 2013a), mobile phone call records (Gonzalez et al., 2008), and taxi trajectories (Peng et al., 2012). Studies (Peng et al., 2012) reveal that the probability of visited locations follows a Zipf's law. Using smart card data, Hasan et al. (Hasan et al., 2013a) found that the two most visited places have similar probabilities and the probability distribution of the places with rank greater than 2 follows a Zipf's law. When it comes to ride-hailing service users, we find the similarity of the probabilities of the two most visited places, similar to the results found by Hasan et al. (Hasan et al., 2013a). Additionally, the distribution of the visited place rank probabilities follows a Zipf's law when the number of visited place is higher.

Additionally, we visualize the spatial distribution of home and work places of on-demand ride-hailing service users. Compared with the public transit users (Ma et al., 2017), jobs-housings of on-demand ride-hailing users does not show severe imbalance possibly because people are less likely to take ride-hailing services for long-distance commuting purposes. We also present the travel distance distribution in spatial scale for commuting and non-commuting trips to validate that on-demand users prefer to commute for short distance.

Finally, we have reported the distributions of travel distance and travel time both for commuting trips and non-commuting trips of ride-hailing service users, capturing the patterns of another significant aspect of spatial regularity. The travel distance and travel time for both commuting and non-commuting trips show similar patterns observed in other studies (Zheng et al., 2018, Zhao et al., 2015). It is worth mentioning that the average travel distance and travel time distribution of commuting trips presents more left-skewed compared to that of non-commuting trips. It implies that people tend to travel less distances when they commute by a ride-hailing service, probably due to economic considerations. Distances of commuting and non-commuting trips follow a power log-normal distribution, while previous studies (Zhao et al., 2015) found that the distances of taxi trips follow a log-normal distribution. In addition, we also fit the distributions of travel times of commuting and non-commuting trips. The travel time of commuting trips follows an exponential Weibull distribution and the travel time of non-commuting trips follows a log-normal distribution.

**IMPLICATIONS**

The results of our analysis provide several implications for traffic management and urban planning, which are summarized as follows:

1. Urban land use characteristics have significant influence on individual commuting patterns (Suzuki and Lee, 2012). Thus, understanding the spatial distributions of individual residences and work places are the key points for urban planning and policy decisions (Aguilera and Voisin, 2014). In this paper, we provide a cost-effective way to collect large-scale data on individual home and workplaces based on the emerging ride-hailing trip data.

2. Predicting traffic states is one of the critical aspects of deploying intelligent transportation systems. Accurate and reliable prediction of travel states such as travel time enables people to make informed travel decisions. Individual mobility patterns presented in this study will help to develop high-resolution individual-level demand prediction models that can be applied to improve the performance of transportation network (Wen et al., 2019).

3. The distribution of trip generation, gap time, travel distance and travel time can be used in agent-based traffic simulations, which is one of the most essential tools for evaluating the expected performance of a new policy as well as designing innovative services using emerging technologies such as the connected and autonomous vehicles and mobility as a service (Wen et al., 2019).

4. Ride sharing is considered to be one of the potential ways to solve the congestion problems and meet the commute demand, especially with autonomous vehicles since it can increase automobile occupancy (Lavieri and Bhat, 2019). Individual mobility behaviors will provide useful information to develop matching algorithms for potential passengers and drivers, which can significantly improve ride sharing services.

**CONCLUSIONS**

In this paper, we reveal the spatio-temporal patterns of individual mobility from the perspective of ride-hailing service passengers. The empirical analysis of massive movement data provides us deeper insights on individual mobility patterns at a city level. Regarding temporal movement patterns, we capture the difference of trip generation characteristics between weekdays and weekends and the distribution of gap time between consecutive trips. In terms of spatial mobility patterns, we visualize the distribution of home and work places as well as the travel distance in spatial scale and observe the distribution of the number of visited place and their rank and report the distribution of travel distance and travel time.

The emergence of ride-hailing services can help serve the growing transportation demand of our expanding cities, significantly improving the quality of city life and access to different places. From a spatio-temporal perspective, the study findings help us better understand human movement patterns. This study provides new insights on modeling travel behavior of ride-hailing service users. It shows the potential to predict individual movement using this emerging mode. Our results also provide insights to develop high-fidelity simulations of on-demand service operations, which can further benefit developing services that depend on ride-hailing. In future research, we will focus on developing high-resolution generative models to forecast individual movements in cities.

**Appendix**

In this paper, to fit the empirical data of travel distance and travel time of on-demand ride-hailing users, we apply 6 commonly used distributions: *log-normal, Weibull, Gamma, student's t, exponentiated Weibull* and *power log-normal*. The detail information of the distributions can be seen as follows:

(1) Log-normal:

$$f(x) = \frac{1}{\sigma x \sqrt{2\pi}} \exp\left(-\frac{(lnx - \mu)^2}{2\sigma^2}\right) \quad (1)$$

Where $\mu$ and $\sigma$ represent the mean and the standard deviation of the natural logarithm of the variable.

(2) Weibull:

$$f(x) = \frac{k}{\lambda}\left(\frac{x}{\lambda}\right)^{k-1} e^{-\left(\frac{x}{\lambda}\right)^k} \quad (x \geq 0) \quad (2)$$

Where $k > 0$ is the shape parameter and $\lambda > 0$ is the scale parameter of the distribution.

(3) Gamma:

$$f(x) = \frac{\beta^\alpha x^{\alpha-1} e^{-\beta x}}{\Gamma(\alpha)} \quad (x > 0) \quad (3)$$

Where $\alpha > 0$ is the shape parameter and $\beta > 0$ is the rate parameter. And $\Gamma(\alpha)$ is the gamma function $\Gamma(\alpha) = (\alpha - 1)!$.

(4) Student's t

$$f(x) = \frac{\Gamma\left(\frac{v+1}{2}\right)}{\sqrt{v\pi}\,\Gamma\left(\frac{v}{2}\right)} \left(1 + \frac{t^2}{v}\right)^{-\frac{v+1}{2}} \quad (4)$$

Where $v > 0$ is the number of degrees of freedom and $\Gamma$ is the gamma function, which can be seen in (3).

(5) Exponentiated Weibull

$$f(x) = \alpha \frac{k}{\lambda} \left(\frac{x}{\lambda}\right)^{k-1} (1 - e^{-(x/\lambda)^k})^{\alpha-1} e^{-(x/\lambda)^k} \quad (5)$$

Where $k > 0$ is the first shape parameter, $\alpha > 0$ is the second shape parameter and $\lambda > 0$ is the scale parameter of the distribution. In particular, there are two special cases: when $\alpha = 1$, this function will be Weibull distribution; when $k = 1$, this function will be exponentiated exponential distribution.

(6) Power log-normal

$$f(x) = (\frac{p}{x\sigma})\phi(\frac{logx}{\sigma})(\Phi(\frac{-logx}{\sigma}))^{p-1} \quad (6)$$

Where $p$ (also referred to as the power parameter) and $\sigma$ are the shape parameters, $\phi$ is the PDF of the standard normal distribution and $\Phi$ is the CDF of the standard normal distribution.